\documentstyle[12pt]{article}
\begin{document}
\title{\bf Oscillations of Cumulant Moments - Universality  of
Amplitudes} 
\author{M.Rybczy\'nski$^{1}$,
        G.Wilk$^{2}$\thanks{e-mail:wilk@fuw.edu.pl},
        Z.W\l odarczyk$^{1}$\thanks{e-mail:wlod@pu.kielce.pl},\\
        M.Biyajima$^{3}$\thanks{e-mail:biyajima@azusa.shinshu-u.ac.jp}
        and N.Suzuki$^{4}$\thanks{e-mail:suzuki@matsu.ac.jp}\\
        [2ex]  
 $^1${\it Institute of Physics, Pedagogical University,}\\
     {\it Konopnickiej 15; 25-405 Kielce, Poland}\\  
     [1ex]
 $^2${\it The Andrzej So\l tan Institute for Nuclear Studies,}\\
     {\it Ho\.za 69; 00-689 Warsaw, Poland}\\
     [1ex]
 $^3$ {\it Department of Physics, Shinshu University,}\\
      {\it  Matsumoto 390, Japan} \\ 
      [1ex]
 $^4$ {\it Matsusho Gakuen Junior College,}\\
      {\it Matsumoto 390-12, Japan}\\
 }
\date{\today}
\maketitle 

\newpage
\begin{abstract}
We demonstrate on simple examples that oscillatory behaviour of 
moments of multiplicity distributions $P(n)$ observed in $e^+e^-$
annihilations, in hadronic $pp$ collisions and in collisions on 
nuclei, $p+A$, is to a large extend caused by the experimental
artifact of measuring only limited range of $P(n)$. In particular we
show that by applying a suitable universal cutt-off procedure to the
measured $P(n)$ one gets for reactions mentioned before oscillations
of similar magnitude. The location of zeros of oscillations as 
a function of the rank of moments and their shapes remain, however, 
distinctively different for different types of reactions considered.
This applies to some extend also to collisions of nuclei, which
otherwise follow their own pattern of behaviour.\\ 


\end{abstract}

\newpage

The problem of the possible physical origin and information content
of oscillations in the cumulant moments of the corresponding
multiplicity distributions $P(n)$ started with QCD calculation of
the respective generating functional. It turned out that the
resulting cumulant moments oscillate as a function of their rank in
the way depending on the QCD parameters used \cite{drem93}. This
finding was confirmed by the analysis of $e^+e^-$ and hadronic $pp$
data, which showed that, indeed, the $q$-th rank normalized cumulant
moment of observed negatively charged multiplicity distributions
oscillates irregularly around zero with increasing $q$ (the minimum
points being located around $q\simeq 5$). There was therefore a hope
that analysis of these oscillations can then prove a crucial test for
the QCD \cite{drem94}.\\    

These expectations were, however, soon confronted with observations
that the same data can be equally well described by a more
phenomenological methods \cite{suzu96,naka96} based on the 
solutions of stochastic processes \cite{biya84}, for example by the
negative binomial distribution (NBD) \cite{suzu91} (in its truncated
version) or by the modified negative binomial distribution (MNBD)
\cite{suzu91,chli90}. The interesting finding was that the NBD and
MNBD differ distinctively in this context in the following sense: for
untruncated multiplicity distributions the $q$-th rank normalized
cumulant moment of the NBD is always positive and decreases
monotically with increasing $q$, whereas for MNBD it can oscillate in
the way depending on the choice of parameters. Therefore, in this
approach, the behavior of cumulant moments obtained from experimental
data seemed to provide a new constraint on models of multiplicity
distributions. In particular it was shown in \cite{suzu96,naka96}
that cumulant moments of negatively charged particles in $e^+e^-$
collisions can be described by both the truncated NBD and (truncated
or not) MNBD (which performs better in the $e^+e^-$ case). \\ 

The observation that truncation of the NBD makes the corresponding
moments oscillate has confirmed the statement made before in Ref.
\cite{ugoc95}. It was said there that important (if not exclusive)
factor leading to oscillations of moments is the experimental fact of
necessary truncation of the observed $P(n)$ at some maximal
multiplicity $n_{max}$. The observed differences between results from
different reactions seem to reflect therefore only the level of this
truncation.\\ 

Cumulant moments obtained from the $hA$ experimental data also
oscillate with magnitude which is much bigger than that observed in
$hh$ collisions \cite{drem97}. The specific feature of these
oscillation is that they can be atributed not only to truncation of
$P(n)$ but also to the fact that the number of elementary collisions
in the $hA$ reaction is necessary limited by the geometry of
collision \cite{suzu98}. This is, in fact, a kind of truncation as
well, but this time it is caused by the geometry of the collision
rather than by the experimental setup. This new geometrical factor
should be therefore even more important in heavy ion collisions
\cite{GEO,ddd97}.\\  

The above discussion clearly shows that information content of
the oscillation phenomenon remains still unclear and subject to
debate. The aim of our note is thus to shed a new light on this
problem by discussing a couple of simple but illustrative numerical
examples of oscillations in $e^+e^-$ annihilations, hadronic $pp$
collisions and collisions involving nuclei, $pA$. We shall not attempt
here a fit to experimental data because this was done already in the
relevant works quoted here. Our intention was rather to use the existing
experience on this subject (especially that contained in
\cite{suzu96,naka96,suzu98}) in order to demonstrate a possible
universality existing in the $e^+e^-$, $pp$ and $pA$ data on
oscillations of moments. We shall also address, albeit only shortly,
heavy ion collisions, $AB$, in this context.\\

We shall use, as is usually done, the following moments of the
multiplicity distribution $P(n)$ (cf. \cite{suzu98}):    
\begin{equation}
H_q\, =\, \frac{K_q}{F_q} \label{eq:Hq}
\end{equation}
where 
\begin{equation}
K_q\, =\, \frac{k_q}{\langle n\rangle ^q}, \qquad 
F_q\, =\, \frac{f_q}{\langle n\rangle ^q} \label{eq:KF}
\end{equation}
with $k_q$ and $f_q$ being the usual cumulant and factorial moments
of rank $q$ of $P(n)$. What is observed experimentally is the fact
that they oscillate and that these oscillations differ substantially
depending on the type of reaction and do it in two ways: 
\begin{itemize}
\item their amplitudes vary, increasing from the value of $10^{-3}$
for $e^+e^-$ annihilations, via the value of $10^{-2}$ for $pp$
collisions up to the value of $10^{-1}$ for $pA$ and heavy ion ($AB$)
reactions;
\item their shapes are different with frequency of oscillations in
$q$ being highest for $e^+e^-$ reactions.
\end{itemize}
As was said before, the main cause of these oscillation is supposed
to be experimental truncation of the corresponding $P(n)$. In order
to compare results of such truncation for different reactions we
propose to use a universal variable $u$ defined in the following way:
\begin{equation}
u\, =\, \frac{n_{max}\, -\, \langle n\rangle}{\sigma_n}. \label{eq:U}
\end{equation}
This variable measures distance of the cut-off point, $n_{max}$, from
the (specific for the process under consideration) mean multiplicity
$\langle n\rangle$. It does this in terms of standard deviation
$\sigma_n$ (which is obtained from the same $P(n)$). In this way it
allows to compare results from different reactions by providing a
kind of natural and universal measure for terminating $P(n)$ under
considerations at some value of $n_{max}$.\\ 

In Fig. 1 we show examples of $H_q$ moments calculated for $e^+e^-$,
$pp$ and $pA$ reactions for two different choices of the values of
cut-off parameter $u$. In each case we have used identical
multiplicity distributions (and all other relevant parameters) as 
those used in Refs. \cite{suzu96,naka96,suzu98} when describing the same
reactions. They were then cut-off for each reaction considered at the
same value of the variable $u$ defined in eq. (\ref{eq:U}) and from
them the corresponding moments $H_q$ were calculated. As can be seen,
cutting off multiplicity distributions $P(n)$ (calculated for
different reactions) at the same values of $u$ results in comparable 
values of amplitudes of observed oscillations. Although they are
still not identical, the previously mentioned differences in
amplitudes are enormously reduced, being now of the same order of 
magnitude. This feature apparently does not depend on the actual
value of variable $u$ used (although the changes of $u$ affect the
shape of oscillations). It proofs therefore that the increase of
amplitudes of oscillations observed between $e^+e^-$ and $pA$
reactions is caused mainly by different experimental cut-off
procedures (quantified here by different values of variable $u$ used
in the respective processes) applied to the measured $P(n)$. In
$e^+e^-$ processes, with smallest amplitudes of oscillations, the
$P(n)$ were measured most accurately, up to the very high
multiplicities (i.e., to large values of the ratio $z=n/\langle
n\rangle$). The opposite situation is encountered in $pA$ processes.
This is the main result of our note.\\    

This kind of universality (even if only approximate) makes the sizes
of amplitudes of oscillations not particularly sensitive to the
dynamical details of $P(n)$ of interest. Not much is left in this
observable when different experiments, but with the same values of
variable $u$, are compared with each other. On the other hand, the
character of oscillations, as visualised by their frequency in the
rank $q$ of moments, remains in a visible way different for different
types of reactions and can therefore be used for dynamical 
discrimanation between different models. For example, in Fig. 2 we
show oscillations of $H_q$ moments obtained for the same value of
$u=7$ using $P(n)$ in the form of MNBD as given in
\cite{suzu96,naka96}:    
\begin{eqnarray}
P(0)\, &=&\, \frac{\left( 1 + r_1\right)^N}{\left( 1 +
r_2\right)^{N+k}} ,\nonumber\\
P(n)\, &=&\, \frac{1}{n!}\, \left(\frac{r_1}{r_2}\right)\,
\sum_{j=0}^N\, _NC_j\, \frac{\Gamma(k+n+j)}{\Gamma(k+j)}\, 
\left(\frac{r_2 - r_1}{r_1}\right)^j \, 
\frac{r_2^n}{\left( 1 + r_2\right)^{n+k+j}} , \label{eq:MNBD}
\end{eqnarray}      
where $N,~k,~r_{1,2}$ are parameters. Referring to
\cite{suzu96,naka96} for details we shall say only that if $k=0$, the
summation in eq. (\ref{eq:MNBD}) runs from $j=1$ up to $j=N$ and the
resultant distribution is called MNBD. In this case parameters
$r_{1,2}$ are given by the average multiplicity $\langle n\rangle$ and
second moment $C_2$ of the corresponding $P(n)$ as
$r_{1,2}=\frac{1}{2}\left(C_2 - 1 -\frac{1}{\langle n\rangle} \mp
\frac{1}{N}\right)\langle n\rangle$. For $N=0$, parameter $r_1$
disapears from (\ref{eq:MNBD}) and it reduces to the NBD with $r_2 =
\frac{\langle n\rangle}{k}$. The parameters $r_1$ and $r_2$ of the
NMBD reflect now the structure of oscillations rather than their
amplitudes. As one can see they change systematically from parameters
describing $e^+e^-$ annihilation (left-top panel) to those typical
for hadronic $pp$ collisions (right-bottom panel)
\cite{suzu96,naka96}.\\       

To summarize, we stress again that the magnitude of observed
oscillations of $H_q$ moments of multiplicity distributions $P(n)$
reflect essentially our ability to measure, in a given reaction,
large multiplicities. When analysing data using the same value of our
universal cut-off parameter $u$ one gets comparable values of
amplitudes for all reactions of interest. It means that this quantity
is not sensitive to dynamical details of reaction. The shape of
oscillations remains, however, sensitive to such details. It can
therefore be used to extract a new dynamical information from
different multiplicity distributions (when compared at the same
values of the cut-off parameter $u$).\\

The separate issue is the problem of oscillations in heavy ion 
collisions $AB$, which we should now briefly address for the sake of
the completeness of presentation. They do not share the property
discussed above. The reason is the following. As was already
mentioned, in the collisions of two nuclei, $A$ and $B$, the nuclear
geometry is the main factor responsible for the shape and properties
of the corresponding multiparticle distribution of produced
secondaries $P(n)$ \cite{GEO}. This fact is crucial in generating
oscillations in the respective cumulant moments. To show it on some
example let us first write the typical corresponding multiplicity
distribution for $A+B$ collision: 
\begin{equation}
P(n)\, =\, \sum_{\mu=1}^{\mu_{tot}}\, p(\mu)\, \prod_{i=1}^{\mu}\, P_i(n_i)\,
       \delta\left(n\, -\, \sum_{i=1}^{\mu}\, n_i\right). \label{eq:Pn}
\end{equation}
It contains two ingredients: distribution $p(\mu)$ of the number of
emitting sources $\mu$ and the respective ``elementary'' multiplicity
distributions of particles produced from such sources, $P_i(n_i)$.
The emitting sources can be, for example, understood as in
\cite{BSWW}. Their distribution can be calculated in the same way as 
in \cite{suzu98,GEO,ddd97}. In the example below we have used a
simple Monte Carlo code in which two colliding nuclei consisting of 
$A$ and $B$ nucleons, respectively, collide with each other. Nucleons
are distributed in nucleus according to a standard Saxon-Woods (SW)
distribution (with diffusiveness $0.49$ fm for $S$ and $0.545$ fm for
$Pb$ nuclei and corresponding nuclear radii given by the formula:
$r[{\rm fm}] = 1.12 A^{1/3}-0.86 /A^{1/3}$). They collide with each
other with probability given by their (total inelastic) cross section
$\sigma =32$ mb. This provides us with $p(\mu)$. On the other hand
$P_i(n_i)$ has been taken again from the MNBD fits to elementary
collisions performed in \cite{suzu96,naka96}). In nuclear collisions
two distinct classes of events occur and must be treated separately:
central and minimum bias collisions. In our case central collisions
were chosen as $1\%$ of the collisions with smallest impact
parameter. In Fig. 3 we show results for moments $H_q$ of $P(n)$ from
eq. (\ref{eq:Pn}) calculated for $S+S$ (left panels) and $Pb+Pb$
(right panels) minimum bias (upper panels) and central (lower panels)
collision for the same values of the variable $u=5$. Notice that the
magnitude of amplitude of oscillations (especially for central
collisions) remains different from that in the corresponding panels
of Fig. 1. It means that in this case there is no such universality
as in the previously discussed reactions. On the other hand, however,
minimum bias collisions are distinctively different from the central
ones, which  show only very small oscillations. The patterns shown
apparently depend only weakly on the choice of the colliding nuclei
(i.e., on the parameters of the Monte Carlo producing $p(\mu)$).\\

To understand better results presented in Fig. 3 one should realise
that central collisions result in a large number of elementary
collisions, i.e., in a large number of emitting sources $\mu$ in each
event. Therefore, because of central limit theorem, irrespectively of 
details of elementary collisions $P(n)$ must have a gaussian-like
shape. We can parametrize it as: $P(n) = P_0\cdot \exp\left( -
\frac{(n - n_0)^2}{2a}\right)$. On the other hand, the minimum bias
collisions result in $p(\mu)$ of the box-like, or Saxon-Woods
(SW)-like shape and such will be also resultant $P(n)$: $P(n) =
\frac{P_0}{1\, +\, e^{(n - n_0)/a}}$. In Fig. 4 we show, as
illustration, some typical examples of oscillation patterns emerging
from both types of distributions. Notice that whereas we essentially
observe no oscillations in the case of gaussian $P(n)$ (or, if at
all, they do start at large $q$), we see strong oscillations for the
SW $P(n)$. They are caused in this case by the box-like shape of the
SW distribution, which is best demonstrated by the fact that they
gradually vanish with the increasing difussiveness of SW used, i.e.,
with the increasing values of parameter $a$ \cite{FOOT1}. It should
be pointed here that results presented in Fig. 4 were obtained
without additional truncation in multiplicity, i.e., in (\ref{eq:U})
$n_{tot} = \infty$. All oscillations present there are thus entirely
of different origin than the simple truncation of $P(n)$. They are
governed by the geometrical parameter $a$ and by the level of
observability of the total $P(n)$.\\

Summarizing, we have demonstrated (approximate) universality of
amplitudes of oscillations of cumulant moments when compared at the
same values of the variable $u$ as defined in (\ref{eq:U}). It shows
up for a range of reactions from $e^+e^-$ annihilation processes, via
hadronic $pp$ collisions, to $pA$ reactions. The latter start to show
influence of the geometry of collision process, which entirely
dominates the truly nuclear collisions.\\

\vspace{3mm}
\noindent
{\bf Acknowledgements}\\
G.W. would like to extend his gratitude to the Physic Department 
of Shinshu University and to Matsusho Gakuen Junior College for
their warm hospitality during his visit to Matsumoto where this work
originated. M.B. is partially supported by the Grant-in Aid
for Scientific Research from the Ministry of Education, Science
and Culture (No. 06640383) and by the Exchange Program between JSPS
and the Polish Academy of Science.  N.S. thanks for the financial
support by Matsusho Gakuen Junior College.\\

\vspace{3mm}

%
\newpage
\noindent
{\bf Figure captions}\\
\begin{itemize}

\item[{\bf Fig. 1}]  Examples of $H_q$ moments of multiplicity
                     distributions for the $e^+e^-$ annihilation
                     (upper panels), $pp$ reactions (middle pannels)
                     and $pA$ reactions (bottom panels) for two
                     chosen values of the parameter $u$: $u=5$ (left
                     panels) and $u=8$ (right pannels). (The $P(n)$
                     data are the same as in 
                     \cite{drem93,drem94,suzu96,naka96,suzu98}).

\item[{\bf Fig. 2}] The $H_q$ moments obtained from the MNBD for $P(n)$
                    for different values of its characteristic parameters
                    $r_1$ and $r_2$ (cf. \cite{suzu96,naka96}). The 
                    upper-left panel corresponds to $e^+e^-$ and bottom-right
                    one to $pp$ reactions, respectively. The value of 
                    parameter $u=7$ reamains all time the same. Notice 
                    the gradual change of frequency of oscillations 
                    whereas their amplitudes remain essentially of the 
                    same order of magnitude.
                    
\item[{\bf Fig. 3}] The $H_q$ moments calculated for 
                    $S+S$ (left panels) and $Pb+Pb$ (right panels)
                    collisions of the minimum bias (upper panels)
                    and central (lower panels) type. In both cases
                    $u=5$.

\item[{\bf Fig. 4}] Examples of oscillation patterns for gaussian-like 
                    (upper panels) and SW-like (lower panels) shapes
                    of multiplicity distributions $P(n)$, see text for
                    details. Short dash, long dash and full lines 
                    correspond to parameter $a$ equal to 
                    $a = 2,~20~,80$ for gaussian-like
                    distributions and to $a = 99,~50,~10,~0.001$ 
                    for SW-like distributions; in both cases $n_0 = 400$.

\end{itemize}
\newpage
\centerline{Figure 1}
\begin{figure}[h]
\setlength{\unitlength}{1cm}
\begin{picture}(15.,10.)
\includegraphics{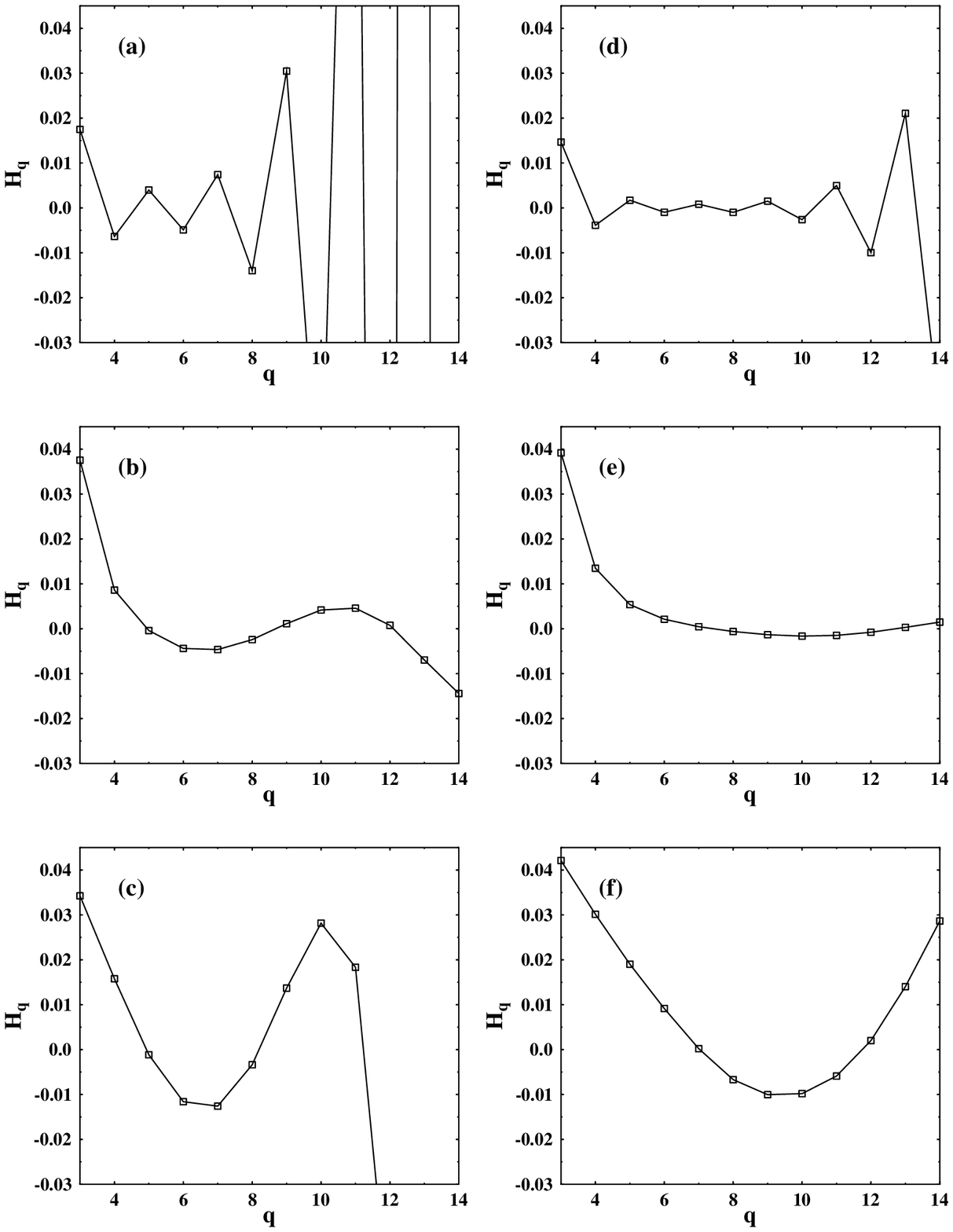}
\end{picture}
\end{figure}
\newpage
\centerline{Figure 2}
\begin{figure}[h]
\setlength{\unitlength}{1cm}
\begin{picture}(15.,10.)
\includegraphics{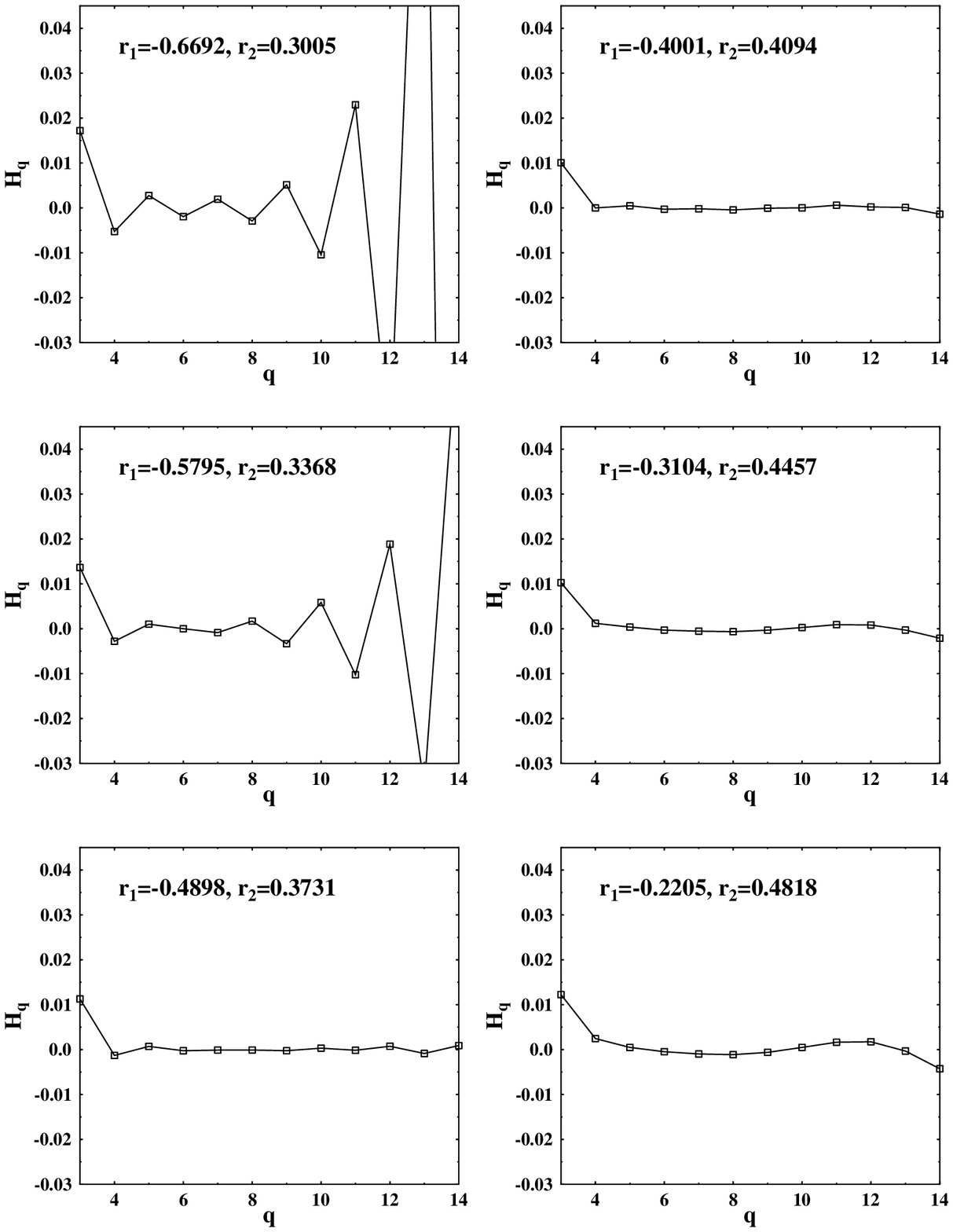}
\end{picture}
\end{figure}
\newpage
\centerline{Fig. 3}
\begin{figure}[h]
\setlength{\unitlength}{1cm}
\begin{picture}(15.,10.)
\includegraphics{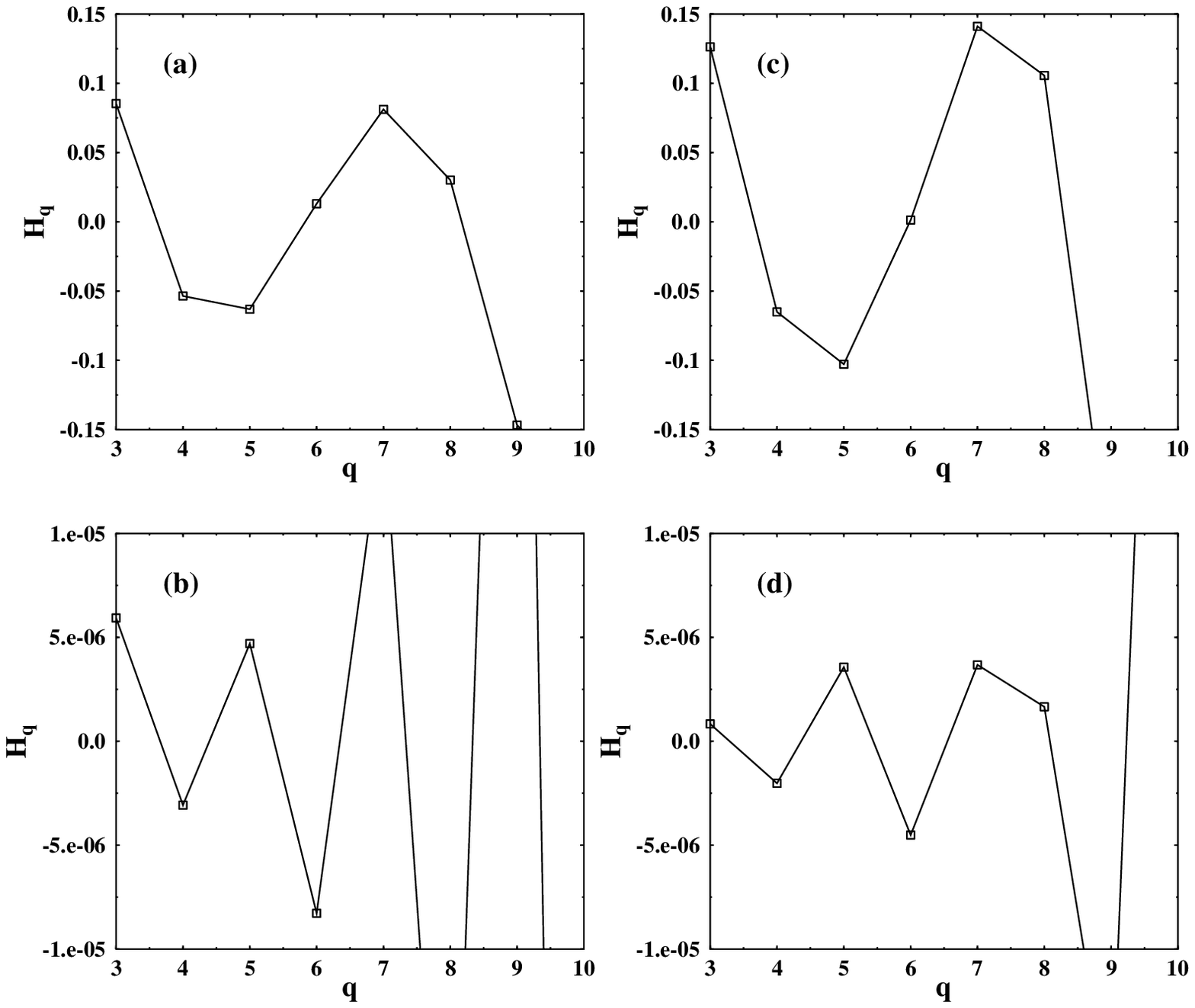}
\end{picture}
\end{figure}
\newpage
\centerline{Fig. 4}
\begin{figure}[h]
\setlength{\unitlength}{1cm}
\begin{picture}(15.,10.)
\includegraphics{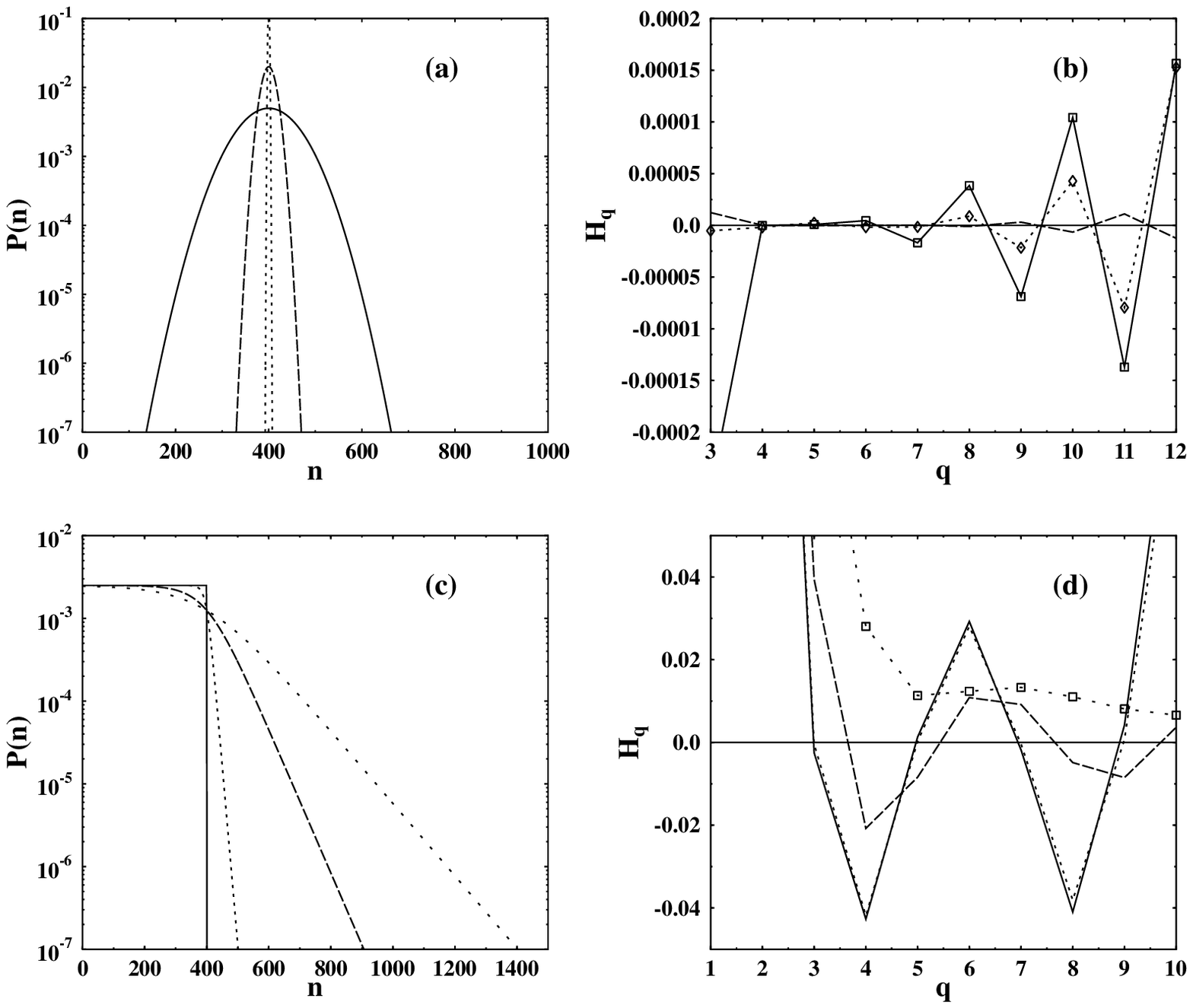}
\end{picture}
\end{figure}

\begin{thebibliography}{cc}
%
 \bibitem{drem93} I. M. Dremin and V. A. Nechitailo, {\sl JETP
                  Lett.} {\bf 58} (1993) 881; \\
                  I. M. Dremin and R. Hwa, {\sl Phys. Rev.} {\bf
                  D49} (1994) 5805.
%
 \bibitem{drem94} I. M. Dremin, et al. {\sl Phys. Lett.} {\bf B336}
                  (1994) 119.
%
 \bibitem{suzu96} N. Suzuki, M. Biyajima and N. Nakajima,
                  {\sl Phys. Rev.} {\bf D53} (1996) 3582 and 
                  {\bf D54} (1996) 3653.
%
 \bibitem{naka96} N. Nakajima, M. Biyajima and N. Suzuki,
                  {\sl Phys. Rev.} {\bf D54} (1996) 4333.
%
 \bibitem{biya84} M. Biyajima and N. Suzuki, {\sl Phys. Lett.}
                  {\bf 143B} (1984) 463 and 
                  {\sl Prog. Theor. Phys.} {\bf 73} (1985) 918.
%
 \bibitem{suzu91} N. Suzuki, M. Biyajima and G. Wilk,
                  {\sl Phys. Lett.} {\bf B268} (1991) 447.
%
 \bibitem{chli90} P. V. Chliapnikov and O. G. Tchikilev,
                  {\sl Phys. Lett.} {\bf B242} (1990) 275 and
                  {\bf B282} (1992) 471; \\
                  P. V. Chliapnikov, O. G. Tchikilev and V. A.
                  Uvarov, {\sl Phys. Lett.} {\bf B352} (1995) 461.
%
 \bibitem{ugoc95} R. Ugoccioni, A. Giovannini and S. Lupia,
                  {\sl Phys. Lett.} {\bf B342} (1995) 387.
%
 \bibitem{drem97} A.Capella, I.M.Dremin, V.A.Nechitailo and J.Tran
                  Than Van, {\sl Z.Phys.} {\bf C75} (1997) 89;\\
                  I. M. Dremin, V. A. Nechitailo, M. Biyajima
                  and N. Suzuki, {\sl Phys. Lett.} {\bf B403} (1997)
                  149.
%
 \bibitem{suzu98} N.Suzuki, M.Biyajima, G.Wilk and Z.W\l odarczyk,
                  {\sl Phys. Rev.} {\bf C58} (1998) 1720.
%
 \bibitem{GEO} W.Q.Chao and B.Liu, {\sl Z. Phys.} {\bf C42} (1989)
               337; P.Zhuang and L.Liu, {\sl Phys. Rev.} {\bf D42}
               (1990) 848.
%
 \bibitem{ddd97} J.Dias de Deus, C.Pajares and C.A.Salgado,
                 {\sl Phys. Lett.} {\bf B407} (1997) 335.
%
 \bibitem{BSWW} M. Biyajima, N. Suzuki, G. Wilk and Z. W\l odarczyk,
                {\sl Phys. Lett.} {\bf B386} (1996) 297.
%
 \bibitem{FOOT1} It is interesting to mention that making 
                 $p(\mu)$ more diffuse (similar to $a=99$
                 case in Fig. 4) results in the vanishing
                 of the oscillation pattern (in spite of
                 the fact that MNBD used for elementary
                 $P_i(n_i)$ lead always to oscillations
                 of moments in elementary collisions 
                 \cite{suzu96,naka96}).

%
\end{thebibliography}
\end{document}